\documentclass[review]{elsarticle}

\usepackage{lineno,hyperref}
\usepackage{amssymb}
\usepackage{graphicx}
\usepackage{amsmath}
\usepackage
[
    ,letterpaper
    ,left       = 1in
    ,right      = 1in
    ,top        = 1.5in
    ,headheight = 1in
    ,headsep    = .3in
]{geometry}

\journal{Journal of \LaTeX\ Templates}

\begin{document}

\begin{frontmatter}

\title{Performance Enhancement of Hybrid SWIPT Protocol for Cooperative NOMA Downlink Transmission}
\tnotetext[mytitlenote]{This work was supported by Priority Research Centers Program through the National Research Foundation of Korea (NRF) funded by the Ministry of Education, Science and Technology (2018R1A6A1A03024003)}

\author{A.A.Amin, M.B.Uddin and S.Y.Shin}
\address{}
\fntext[]{}
\author{}
\address{}
\fntext[]{}

\cortext[mycorrespondingauthor]{Soo Young Shin}
\ead{amin@kumoh.ac.kr, ahad.belal@kumoh.ac.kr and wdragon@kumoh.ac.kr}

\address[mymainaddress]{Department of IT Convergence Engineering, Kumoh National Institute of Technology, Gumi, South Korea}
\begin{abstract}

Time splitting and power splitting incorporating, a hybrid Simultaneous Wireless Information and Power Transfer (SWIPT) based cooperative Non-Orthogonal Multiple Access (CNOMA) protocol is considered in this paper. Cell center user of the CNOMA system acts as a relay to enhance the reliability of the cell edge user (CEU). SWIPT is considered to empower the relay operation to avoid the battery draining issue. To enhance the system performance in terms of ergodic sum capacity (ESC) and outage probabilities (OP), an integration of CNOMA strategy and hybrid SWIPT protocol for the downlink (DL) transmission is proposed here. By utilizing the idle link of hybrid SWIPT protocol an enhanced hybrid SWIPT protocol is proposed here to enhance the performance of CNOMA DL transmission.  Moreover, Maximal ratio combining is utilized as a diversity combining technique at CEU to enhance the performance as well. The performance of the proposed protocol is examined in terms of ergodic sum capacity, outage probabilities and energy efficiency. Finally, the analytical results are justified by the Monte-Carlo simulation. Numerical results demonstrate that the proposed protocol with effective CNOMA strategy achieves superior performance than HS-CNOMA with selection combining.
\end{abstract}

\end{frontmatter}

\section{Introduction}
Non-Orthogonal Multiple Access (NOMA) facilitates the simultaneous data transmission to multiple users over the same radio resource by allocating power domain multiplexing [1]. NOMA achieved a great interest in industrial and academic level due to high spectral efficiency and serve a significant amount of users as well [2]. Power domain NOMA and Code domain NOMA are two basic types of NOMA [3]. Since most of the existing researches are based on power based NOMA hence power domain NOMA is considered here [4-5].
\par
In downlink CNOMA, the transmitter transmits a superposed signal to the intended users and each of the users receives that composite signal. Afterward, the cell center user (CCU) performs successive interference cancellation (SIC) to decode its own signal from the received signal [6]. According to the theory of NOMA, the signal power for CEU is significantly higher than CCU hence this is directly decoded at CEU and considering the signal of CCU as noise. Moreover, Cooperative non-orthogonal multiple access (CNOMA) improves reliability and coverage area of a wireless communication system by performing user as a relay or utilizes dedicated relay [7-8]. In this paper, CCU is considered as a relay to enhance the reliability and coverage area of the cell edge user (CEU)[9]. 
\par
The main challenge is to perform the relaying operation by CCU without draining out the battery of the user equipment [6,9,14]. This situation can turn off CCU equipment and causes the degradation of the performance of the network due to the absence of the relay link. Hybrid Simultaneous Wireless Information and Power Transfer (HS) can empower the relay operation of CCU [10-11] for CNOMA system. In addition, the complex transmits antenna selection (TAS) technique is proposed for Multiple Input Single Output (MISO) based system SWIPT system for CNOMA [10-12]. Due to simplicity and energy efficiency, Single Input Single Output (SISO) is considered here. Moreover, complex TAS is not required for SISO based system. In [10], BS to CEU link was idle for the existing HS protocol. Which degrades the performance of the network. So to overcome this efficiency loss, we have proposed data transmission by utilizing this idle link in the proposed enhanced HS (EHS) protocol. In addition, the outage probability of a hybrid SWIPT based CNOMA protocol is investigated only in [10-11]. However, the ergodic sum capacity (ESC) and energy efficiency (EE) of the system is not explored extensively here and the mathematical analysis is not performed to validate the simulation results for ESC [10]. To enhance the ESC of hybrid SWIPT protocol, an enhanced hybrid SWIPT protocol (EHS) is proposed in this paper with an efficient CNOMA strategy. In this strategy, the idle link from the base station (BS) to CEU can be utilized to transfer additional symbol to the CEU. As a result, the ESC is enhanced for the DL transmission of the SWIPT protocol. Moreover, EE is also analyzed in this paper for the proposed EHS and conventional HS protocol for CNOMA as well which was not explored in previous literature.   
\par
In [10], the selection combining (SC) technique is used at CEU as a combining technique. Maximal ratio combining (MRC) is superior to SC as an effective combining technique for information decoding (ID) [15]. That's why MRC is considered here at CEU as combining technique. The contributions of this letter are enumerated below:
\begin{itemize}
\item EHS protocol is proposed and integrated with CNOMA downlink (DL) transmission for two users. MRC is considered at CEU to combine direct and relay link and perform ID effectively. 
\item The ergodic sum capacity (ESC) and outage probabilities (OP) of each user for EHS with CNOMA are investigated and analytical derivations are performed for the Rayleigh fading channel.  

\item The EE is also analyzed for the EHS-CNOMA and HS-CNOMA to evaluate the performance of the proposed EHS protocol over conventional HS protocol. 

\item By using simulation and analysis, the performance improvement of the proposed EHS protocol with considered SISO based CNOMA DL transmission (EHS-CNOMA) over existing HS protocol with SISO based CNOMA DL transmission with SC (HS-CNOMA) are analyzed explicitly.  

\end{itemize}

The rest of the paper is organized as follows: 
Section 2 describes the proposed protocol with system model. Moreover, the system architecture and CNOMA strategy are also described elaborately in this section. Section 3 mathematically analyzes the ergodic sum capacity. Section 4 analyzes the OP for the proposed EHS protocol with the considered model. Section 5 analyzes the ESC for the proposed EHS protocol with the considered model. Section 6 analyzes the EE for the proposed EHS protocol with the considered model. Section 7 exhibits the numerical result analysis. This paper is concluded in Section 8.

\section{Proposed Protocol and System Model}

The two user-based system model is shown in Fig.1. According to the concept of DL CNOMA [4,10-14], the lower power ($p_N$) is assigned for CCU and higher power is assigned ($p_F$) for CEU based on their channel conditions from BS [3]. $d_1$ and $d_2$ are the normalized distances from BS to CCU and CEU respectively. The downlink transmission from BS to the users is performed by two consecutive phases. The transmitted symbols $x_{1}$ and $x_{3}$ are dedicated for CEU and $x_2$ is dedicated for CCU. In (${Ph_1}$), direct signals are transmitted from BS to CCU and CEU respectively.  In second phase ($Ph_2$), a relay link is transmitted from CCU to CEU for $x_3$. Relaying link is required here for $x_3$ at CEU to increase the reliability for $x_3$ at CEU. If the high data rate is required for $x_3$ at CEU, higher signal to noise ratio (SNR) and multiple links can provide the facilities without interruption. That is why a suitable combing technique is considered at CEU for $x_3$. However, relaying link is not considered for $x_1$ or $x_2$. Because $x_1$ is transmitted to CEU with full power so higher reliability can be ensured for $x_1$. Moreover, the additional relaying link for $x_1$ consumes more energy for relaying. In addition, $x_2$ is transmitted to CCU which user has comparatively good channel condition to meet the reliability. SISO based transmitter and receiver is considered in Fig.1.    

All wireless channels are considering as Rayleigh block flat fading channel in this model. Suppose $h_{k}$ denotes the fading coefficient of a channel from BS to a user K, where $K\in(CCU, CEU)$. So $h_{K}$ can be modeled as independent and identically distributed (i.i.d.) complex Gaussian random variables with zero mean and variance $\lambda_{k}$. Moreover, $n_{K}$ denotes additive white Gaussian noise (AWGN) at user K. The variance of the receiver noise is $\sigma_{K}^2$ for the AWGN model.  
\begin{figure}[h]
\centering
\includegraphics[width=0.65\textwidth]{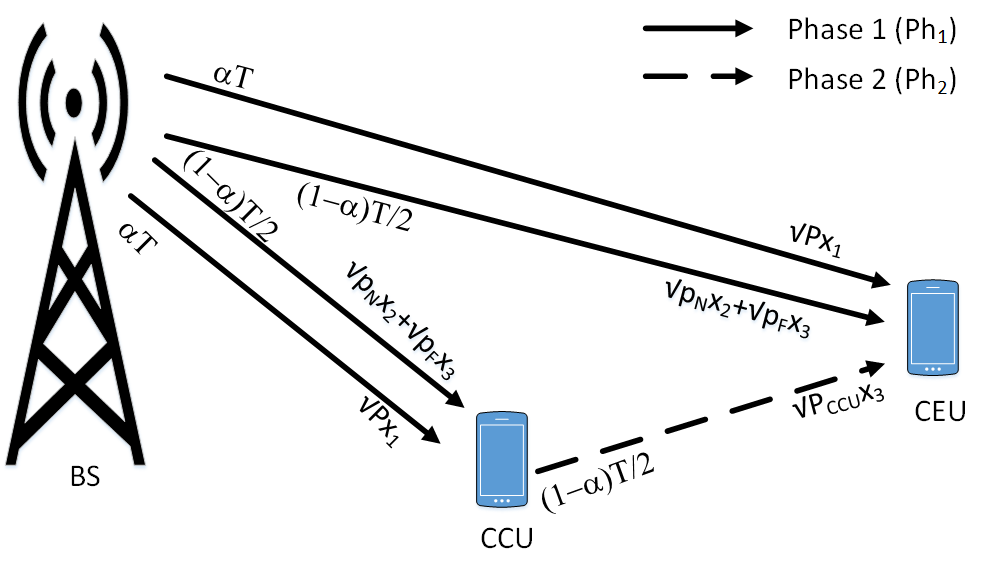}
\caption{System Model} 
\label{image-myimage}
\end{figure}
To energize the relay operation of CCU and information decoding, improved hybrid SWIPT is integrated with the considered model. In addition, hybrid SWIPT is the combination of time switching (TS) and power splitting (PS) methods as well [10-11,18]. Where T is total time duration to perform the DL transmission for CNOMA. So for the TS part, a fraction of time block for energy harvesting is $0<\alpha<1$. Firstly by $\alpha T$ time duration, BS transmits $x_1$ with full power ($P=1$). So, CCU harvested energy from the received observation and CEU receives and decodes $x_1$ at this stage. PS is applied in the ((1-$\alpha$)T/2) duration of time at ${Ph_1}$, the power splitting ratio for PS technique is $0<\delta<1$ and the $\delta$ fraction is utilized for energy harvesting (EH) by PS. Moreover, the $1-\delta$ fraction is used for information decoding (ID) as well at CCU. Furthermore, $x_2$ and $x_3$ symbols are received and decode at CCU and CEU simultaneously from BS. Afterward, CCU exploits all of the harvested energy for information relaying operation by decode and forward (DF) technique [10-11]. The DF is performed at the second phase ($Ph_2$) and ((1-$\alpha$)T/2) duration of time. The EHS protocol for CNOMA is illustrated in Fig.2. 
\begin{figure}[h]
\centering
\includegraphics[width=0.65\textwidth]{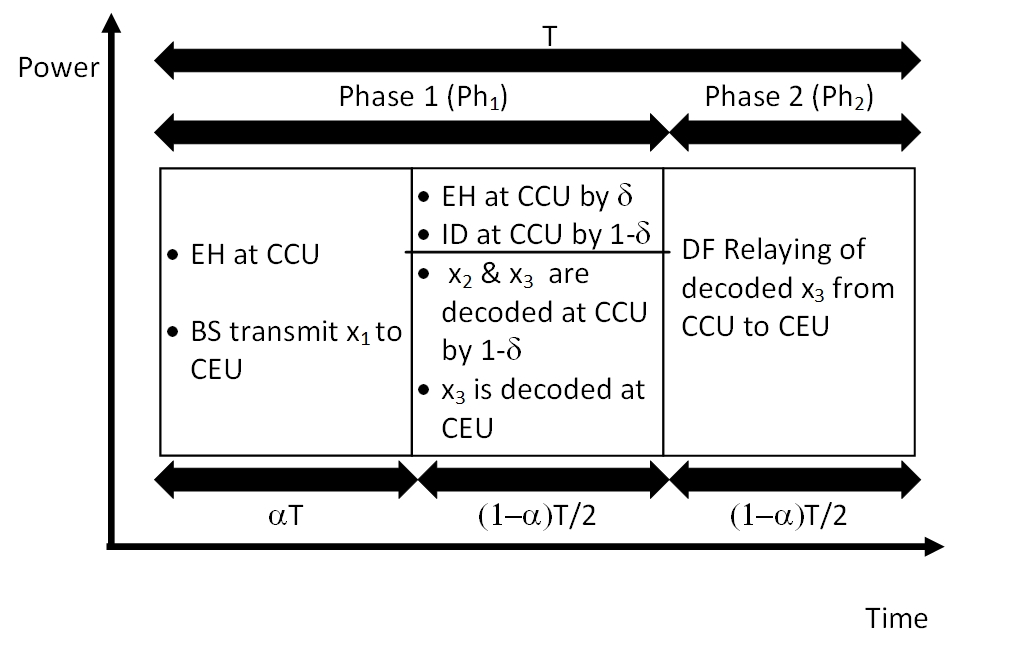}
\caption{Enhanced HS protocol for CNOMA}
\label{image-myimage}
\end{figure}

\subsection{Direct Transmission and Energy Harvesting of Proposed Hybrid SWIPT based CNOMA}
 According to the concept of DL NOMA, the multiplexed signal is directly transmitted to CCU and CEU which illustrates in Fig.1. Where, $0<p_{N}<p_{F}$) and $p_{N} + p_{F} = P$. Here, $P$ is total transmit power. Moreover, $P=1$ is considered here and $\rho$ is the SNR [10,18]. The multiplexed signal is transmitted to CCU and CEU simultaneously. Assume, $|h_{CCU}|^2>|h_{CEU}|^2$, where $|h_{CCU}|^2$ and $|h_{CEU}|^2$ are the channel coefficients for CCU and CEU respectively. The channel fading coefficients between BS and CCU is $h_{CCU}\sim CN(0,\lambda_{CCU})$ with zero mean and variance $\lambda_{CCU}$. Moreover, the Additive White Gaussian noise is $n_{CCU}\sim CN(0,\sigma_{CCU}^2)$ at CCU. Whereas, $\sigma_{CCU}^2$ is the variance of the noise. Similarly, the channel fading coefficients between BS and CEU is $h_{CEU}\sim CN(0,\lambda_{CEU})$ with zero mean and variance $\lambda_{CEU}$. Moreover, the Additive White Gaussian noise is $n_{CEU}\sim CN(0,\sigma_{CEU}^2)$ at the receiver end. $\sigma_{CEU}^2$ is the variance of the noise. In this paper, assume $\sigma_{CCU}^2=\sigma_{CEU}^2=1$ and total time duration is $T=1$ [10-11,18]. At the first phase, energy harvesting is performed by TS at CCU. BS utilizes full power ($p_F=1$) to transmit $x_1$ to CEU. So, the SNR to receive $x_1$ at CEU is expressed as below:

\begin{equation}
\gamma_{CEU}^{x_1}={\rho|h_{CEU}|^2 P}
\end{equation}

By using the concept of DL CNOMA, $x_2$ and $x_3$ are transmitted to CCU and CEU as well at ${Ph_1}$. Successive Interference Cancellation (SIC) receiving technique is considered at CCU in Fig.1. Hence $x_3$ should be decoded first to decode $x_2$ at CCU according to the theory of SIC receiver [6,10-11]. So, the required signal to interference ratio (SINR) to decode $x_3$ at CCU can be expressed as below[10,18]:
\begin{equation}
\gamma_{CCU}^{x_3}=\frac{(1-\delta)p_{F}\rho|h_{CCU}|^2}{p_{N}\rho|h_{CCU}|^2(1-\delta)+(1-\delta)}
\end{equation}

At ${Ph_1}$ the required SNR to retrieve $x_2$ at CCU is expressed as below according to the concept of DL CNOMA [10-11,18]:  

\begin{equation}
\gamma_{CCU}^{x_2}=\frac{(1-\delta)p_{N}\rho|h_{CCU}|^2}{(1-\delta)}
\end{equation}

Moreover, the power of $x_3$ is significantly higher than $x_2$, so $x_3$ can be directly received and decoded at CEU and $x_2$ is treated as noise at CEU and cancel out [11]. In addition, the SNR requirement to receive and decode $x_3$ at CEU is derived as below at ${Ph_1}$[10-11]:   

\begin{equation}
\gamma_{CEU}^{x_3}=\frac{p_{F}\rho|h_{CEU}|^2}{p_{N}\rho|h_{CEU}|^2+1}
\end{equation}

\subsection{Relay Operation by CCU}

CCU utilized the harvested energy to perform relay operation for CEU. So, $|h_{CEU}|^2 <|h_{CCU,CEU}|^2$ should be satisfied to perform the DF relay operation by CCU. In addition, $h_{CCU,CEU}\sim CN(0,\lambda_{CCU,CEU})$ and $h_{CCU,CEU}$ denotes the channel coefficients between CCU and CEU for relay operation. The channel for the relay is modeled as an independent complex Gaussian random variable with zero mean and variance is $\lambda_{CCU, CEU}$. Hence the received signal at CEU from CCU is expressed as below [10-11]:
\begin{equation}
r_{CCU,CEU}=\sqrt{P_{CCU}}x_{3}^{\hat{}} h_{CCU,CEU}+n_{CEU}
\end{equation}
Here, $x_{3}^{\hat{}}$ is the re-encoded version of $x_{F}$. So for $Ph_2$, the required SNR for relay link to receive $x_{3}$ by utilizing the harvested energy is expressed as following equation [7]:
\begin{equation}
\gamma_{Relay}^{x_3}={P_{CCU}|h_{CCU,CEU}|^2} 
\end{equation}
Where, $P_{CCU}=\eta\rho|h_{CCU}|^2|(\frac{2\alpha}{1-\alpha}+\delta)$. $P_{CCU}$ is the transmitted power from CCU to CEU and $\eta$ is energy conversion efficiency. Moreover, $\eta$ plays a vital role to convert the energy from radio frequency signal [7-8]. 
\subsection{Combining Method at Cell Edge User}
Maximal ratio combining (MRC) is considered here at CEU to combine $x_3$ from direct and relay link. Moreover, MRC is used to decode $x_3$ effectively at CEU and increase reliability for the respective symbol [17-18]. Hence, the signal to interference and noise ratio (SINR) for the combined signal is given as below: 
\begin{equation}
  \gamma_{MRC}^{x_3}=\gamma_{CEU}^{x_3}+\gamma_{Relay}^{x_3}   
\end{equation}

\subsection{Channel Capacities}

So from (1) the channel capacity for $x_1$ at CEU is expressed as below [10,18]: 
\begin{equation}
C_{CEU}^{x_1} = \alpha \log_2(1+\gamma_{CEU}^{x_1})
\end{equation}

Consequently, the channel capacity for $x_2$ at CCU from (2) can be derived as following way [10,18]: 

\begin{equation}
C_{CCU}^{x_2} = \frac{(1-\alpha)}{2}\log_2(1+\gamma_{CCU}^{x_2}) 
\end{equation}

Based on the MRC technique at CEU for $x_3$, the channel capacity at CEU is expressed as following equation[10-11,18]: 
\begin{equation}
C_{CEU}^{x_3} = \frac{(1-\alpha)}{2}\log_2(1+\gamma_{MRC}^{x_3}) 
\end{equation}

So the sum channel capacity of the considered system with proposed EHS protocol can be expressed as below[10,18]: 
\begin{equation}
C_{sum}=E[C_{CEU}^{x_1}]+E[C_{CCU}^{x_2}]+E[C_{CEU}^{x_3}]
\end{equation}
Where E[.] is the mean or expectation operation. 

\section{Mathematical Analysis of Ergodic Sum Capacity}

The ESC for EHS-CNOMA with MRC is derived and analyzed in this section. So, from (8) the channel capacity for $x_1$ from BS to CEU for the considered system is written as below [10-11,18]:

\begin{equation}
C_{CEU}^{x_1} = \alpha \log_2(1+{\rho|h_{CEU}|^2}P)
\end{equation}

Let, $X \triangleq {\rho|h_{CEU}|^2 P}$. The CDF of $X$ is written as below: 
\begin{equation}
F_X(x)=1-e^{\frac{-x}{Ps\lambda_{CEU} P}}
\end{equation}

So, the ergodic capacity of $x_1$ at CEU can be achieved by using [13,16] as below: 
\begin{equation}
C_{CEU}^{x_1,erg} =\frac{\alpha}{ln2}(-Ei(-1/g)e^{1/g}) 
\end{equation}

Where, $g=\lambda_{CEU}\rho$ and $Ei[.]$ is the exponential integral. 
Since $x_2$ is the main concern for CCU. So the channel capacity for $x_2$ at CCU for the considered system is expressed as below [10]: 
\begin{equation}
C_{CCU}^{x_2} = \frac{(1-\alpha)}{2}\log_2(1+\frac{(1-\delta)p_{N}\rho|h_{CCU}|^2}{(1-\delta)}) 
\end{equation}
Let $Y \triangleq \frac{(1-\delta)p_{N}\rho|h_{CCU}|^2}{(1-\delta)}$. The CDF of $Y$ can be written as below: 
\begin{equation}
F_Y(y)=1-e^{\frac{-y}{Ps\lambda_{CCU}p_N}}
\end{equation}
So by using [13,16], the ergodic channel capacity of $x_2$ at CCU  can be achieved as below:
\begin{equation}
C_{CCU}^{x_2,erg} =\frac{(1-\alpha)}{2ln2}(-Ei(-1/q)e^{1/q})
\end{equation}
Where, $q=\frac{\lambda_{CCU}\rho(1-\delta)}{1-\delta}$ and $Ei[.]$ is the exponential integral. Furthermore, from (10) the channel capacity at CEU to receive $x_3$ by using MRC technique is given as below:
\begin{equation}
C_{CEU}^{x_3} =  \frac{(1-\alpha)}{2}\log_2(1+\gamma_{MRC}^{x_3}) 
\end{equation}
So by using [17-18], the ergodic channel capacity of (18) can be derived as below:
\begin{equation}
C_{CEU}^{x_3,erg} = \frac{(1-\alpha)}{2ln2}(-Ei(-1/r)e^{1/r}(1+z)
-Ei(-1/s)e^{1/s})
\end{equation}
Where, $r=\eta \rho \lambda_{CEU} (\frac{2\alpha}{1-\alpha}+\delta)$, $z=(p_N/p_F)$, $s= \lambda_{CCU,CEU}$ and $Ei[.]$ is the exponential integral. So the ESC can be achieved by following equation for EHS-CNOMA with MRC: 
\begin{equation}
C_{sum}^{erg}=E[C_{CEU}^{x_1,erg}]+E[C_{CCU}^{x_2,erg}]
+E[C_{CEU}^{x_3,erg}]
\end{equation}
Where $E[.]$ is the mean or expectation operation. 
\section{Outage Analysis}
Let $R_2$ is the predefined data rate of CCU. Moreover, $R_1$ and $R_3$ are the predefined data rate of CEU. The OP of each user is derived as below:

\subsection{Outage Probability of CCU}
The outage event of CCU is occur when CCU cannot decode symbol $x_3$ since SIC is performed at CCU. Moreover, if CCU can decode $x_3$ successfully but cannot decode symbol $x_2$ that can cause outage event as well. Hence OP of CCU can be derived in following way[19-20]: 
\begin{multline} 
OP_{CCU}  =1-Pr{(\gamma_{CCU}^{x_3}>\psi_{CCU}^{R_3})} \times Pr{(\gamma_{CCU}^{x_2}>\psi_{CCU}^{R_2})} \\
          = 1-(1-Pr{(\gamma_{CCU}^{x_3}<\psi_{CCU}^{R_3})}) (1-Pr{(\gamma_{CCU}^{x_2}<\psi_{CCU}^{R_2})}) \\
        =Pr{(\gamma_{CCU}^{x_3}<\psi_{CCU}^{R_3})}+Pr{(\gamma_{CCU}^{x_2}<\psi_{CCU}^{R_2})}-Pr{(\gamma_{CCU}^{x_3}<\psi_{CCU}^{R_3})}Pr{(\gamma_{CCU}^{x_2}<\psi_{CCU}^{R_2})}
\end{multline}
Where, $\psi_{CCU}^{R_2}=2^{(2R_2/(1-\alpha))}-1$ and $\psi_{CCU}^{R_3}=2^{(2R_3/ \\
(1-\alpha))}-1$.Now we can put, 
$Pr{(\gamma_{CCU}^{x_3}<\psi_{CCU}^{R_3})}=\frac{p_{F}\rho\lambda_{CCU}}{p_{F} \\
\rho \lambda_{CCU}+p_{N}\rho\lambda_{CCU}}e^{\frac{- \psi_{CEU}^{R_3}}{\rho \\ \lambda_{CCU}p_F}}$ and $Pr{(\gamma_{x_2}^{CCU}<\psi_{CCU}^{R_2})} = \frac{p_{F}\rho\lambda_{CCU}}{p_{F}\rho\lambda_{CCU}+p_{N}\rho\lambda_{CCU}}(1-e^{\frac{-\psi_{CCU}^{R_2}}{\rho\lambda_{CCU}p_N}})$ into (21) to get the desired OP for CCU as below:
\begin{multline} 
OP_{CCU} =\frac{p_{F}\rho\lambda_{CCU}}{p_{F}\rho\lambda_{CCU}+p_{N}\rho\lambda_{CCU}}e^{\frac{-\psi_{CEU}^{R_3}}{\rho \lambda_{CCU}p_F}}+\\
\frac{p_{F}\rho\lambda_{CCU}}{p_{F}\rho\lambda_{CCU}+p_{N}\rho\lambda_{CCU}}\\
(1-e^{\frac{-\psi_{CCU}^{R_2}}{\rho\lambda_{CCU}p_N}})\\  
 -\frac{p_{F}\rho\lambda_{CCU}}{p_{F}\rho\lambda_{CCU}+p_{N}\rho\lambda_{CCU}}e^{\frac{-\psi_{CEU}^{R_3}}{\rho \lambda_{CCU}p_F}}\\ 
\frac{p_{F}\rho\lambda_{CCU}}{p_{F}\rho\lambda_{CCU}+p_{N}\rho\lambda_{CCU}}(1-e^{\frac{-\psi_{CCU}^{R_2}}{\rho\lambda_{CCU}p_N}})
\end{multline}

\subsection{Outage Probability of CEU}

The OP of CEU can be occur for two different symbols such as $x_1$ and $x_3$. Since $x_1$ and $x_3$ are transmitted from BS to CEU. At ${Ph_1}$, if CEU cannot decode $x_1$ then outage event is occur at CEU for symbol $x_1$. So the OP of CEU for $x_1$ can be derived as below:  

\begin{equation}
OP_{CEU}^{x_1} 
 =Pr{(\gamma_{CEU}^{x_1}<\psi_{CEU}^{R_1})}
\end{equation}

Where, $\psi_{CEU}^{x_1}=2^{(2R_1/(1-\alpha))}-1$ and now we can put $Pr{(\gamma_{CEU}^{x_1}<\psi_{CEU}^{R_1})}=1-e^{\frac{-\psi_{CEU}^{R_1}}{\rho \lambda_{CEU}}}$ in above equation to get the following equation: 

\begin{equation}
OP_{CEU}^{x_1} 
 =1-e^{\frac{-\psi_{CEU}^{R_1}}{\rho \lambda_{CEU}}}
\end{equation}

Moreover, the outage event at CEU for $x_3$ symbol can be happen if CCU or CEU cannot decode $x_3$ successfully. If CCU failed to receive and decode $x_3$ successfully, the outage event is occur at CEU for $x_3$ symbol. In contrast, if CCU can received and decode $x_3$ successfully but $x_3$ cannot decoded successfully by MRC at CEU. This situation also causes outage event at CEU for $x_3$ symbol. So, the outage event occurs at CEU for $x_3$ can be derived as below [19-20]: 
\begin{multline}
OP_{CEU}^{x_3}  =1-Pr{(\gamma_{CCU}^{x_3}>\psi_{CEU}^{R_3})} \times Pr{(\gamma_{MRC}^{x_3}>\psi_{CEU}^{R_3})}\\
        = 1-(1-Pr{(\gamma_{CCU}^{x_3}<\psi_{CEU}^{R_3})}) (1-Pr{(\gamma_{MRC}^{x_3}<\psi_{CEU}^{R_3})})\\
        =Pr{(\gamma_{CCU}^{x_3}<\psi_{CEU}^{R_3})}+Pr{(\gamma_{MRC}^{x_3}<\psi_{CEU}^{R_3})}\\
      -Pr{(\gamma_{CCU}^{x_3}<\psi_{CEU}^{R_3})}Pr{(\gamma_{MRC}^{x_3}<\psi_{CEU}^{R_3})}
\end{multline}

Where, $\psi_{CEU}^{R_3}=2^{(2R_3/(1-\alpha))}-1$ and put  $Pr{(\gamma_{CCU}^{x_3}<\psi_{CEU}^{R_3})}= \frac{p_{F}\rho\lambda_{CEU}}{p_{F}\rho\lambda_{CEU}+p_{N}\rho\lambda_{CEU}}(1-e^{\frac{-\psi_{CEU}^{R_3}}{Ps\lambda_{CEU} p_F}})$ and $Pr{(\gamma_{MRC}^{x_3}<\psi_{CEU}^{R_3})}=1-e^{(\frac{-\psi_{CEU}^{R_3}}{\rho \lambda_{CCU} \lambda_{CCU,CEU}\eta (\frac{2\alpha}{1-\alpha}+\delta)})}$ in (25). So the derived equation after putting the values is like as below:

\begin{multline}
OP_{CEU}^{x_3} 
        =\frac{p_{F}\rho\lambda_{CEU}}{p_{F}\rho\lambda_{CEU}+p_{N}\rho\lambda_{CEU}}(1-e^{\frac{-\psi_{CEU}^{R_3}}{Ps\lambda_{CEU} p_F}})\\
     +(1-e^{(\frac{-\psi_{CEU}^{R_3}}{\rho \lambda_{CCU} \lambda_{CCU,CEU}\eta (\frac{2\alpha}{1-\alpha}+\delta)})})\\
      -\frac{p_{F}\rho\lambda_{CEU}}{p_{F}\rho\lambda_{CEU}+p_{N}\rho\lambda_{CEU}}\\
    (1-e^{\frac{-\psi_{CEU}^{R_3}}{Ps\lambda_{CEU} p_F}})(1-e^{(\frac{-\psi_{CEU}^{R_3}}{\rho \lambda_{CCU} \lambda_{CCU,CEU}\eta (\frac{2\alpha}{1-\alpha}+\delta)})})
\end{multline}

\section{Energy Efficiency}

The harvested energy at CCU by EHS protocol can be derived as below [11],
\begin{equation}
E=\eta \rho |h_{CCU}|^2 \alpha T+\eta \rho |h_{CCU}|^2 (1-\alpha)(T/2)
 \end{equation}
 
In addition, the transmitted power from CCU to CEU can be expressed as below based on the harvested energy $E$ [21],
 
\begin{equation}
P_{CCU}=\frac{E}{(1-\alpha)(T/2)}
 \end{equation}
 
 Moreover, the EE can be derived as below for the proposed EHS-CNOMA scheme [21],
 
\begin{equation}
EE=\frac{C_{sum}}{P_{CCU}}
 \end{equation}
 
 So Eq.29 shows that EE is related to the $C_{sum}$ and $P_{CCU}$.

 \section{Numerical Results and Discussion}
\begin{figure}[h!]
\centering
\includegraphics[width=0.65\textwidth]{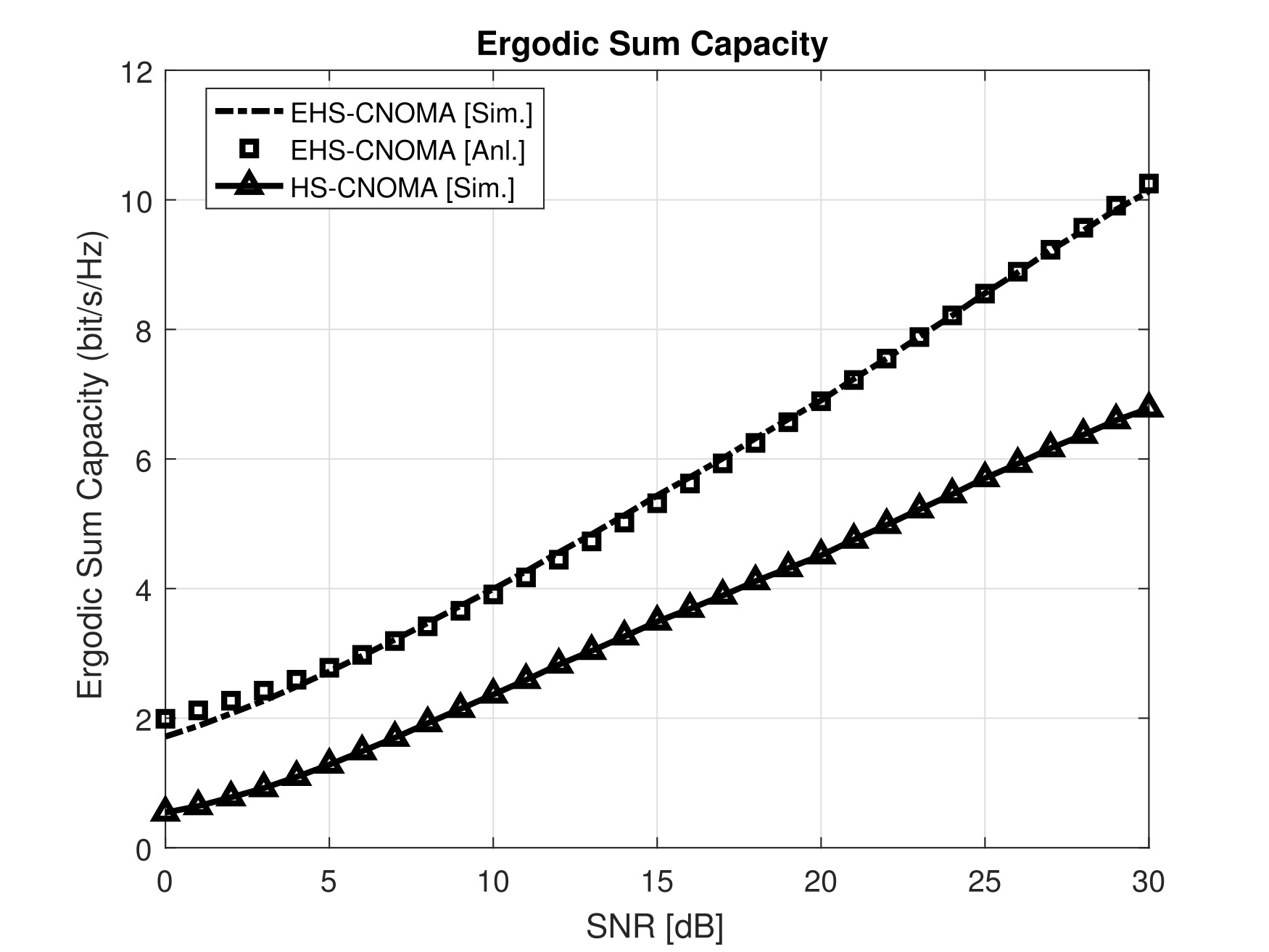}
\caption{ESC versus SNR performance of EHS-CNOMA and HS-CNOMA protocols}
\label{image-myimage}
\end{figure}

\begin{figure}[h!]
\centering
\includegraphics[width=0.65\textwidth]{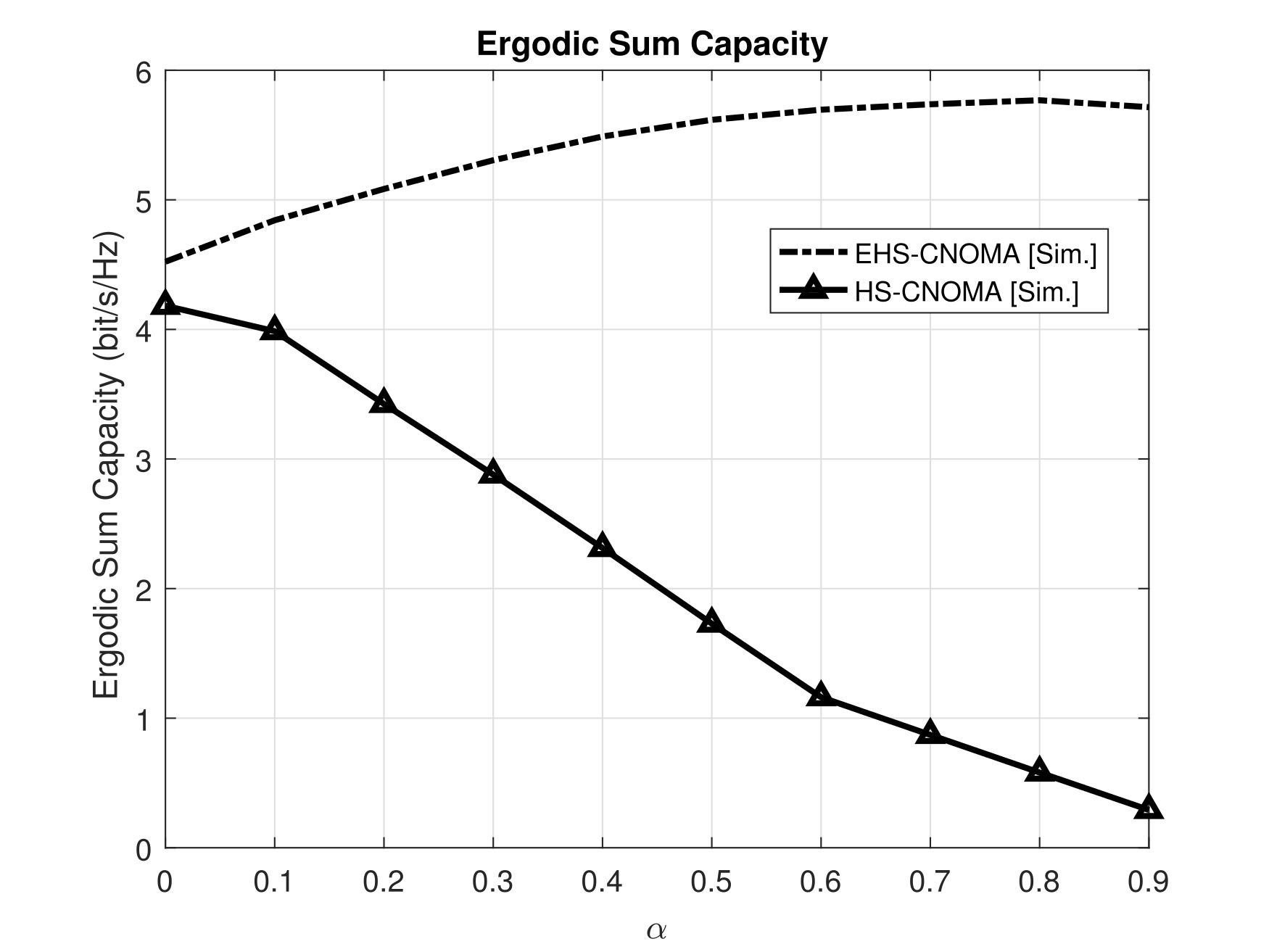}
\caption{ ESC versus $\alpha$ of EHS-CNOMA and HS-CNOMA}
\label{image-myimage}
\end{figure}

\begin{figure}[h!]
\centering
\includegraphics[width=0.65\textwidth]{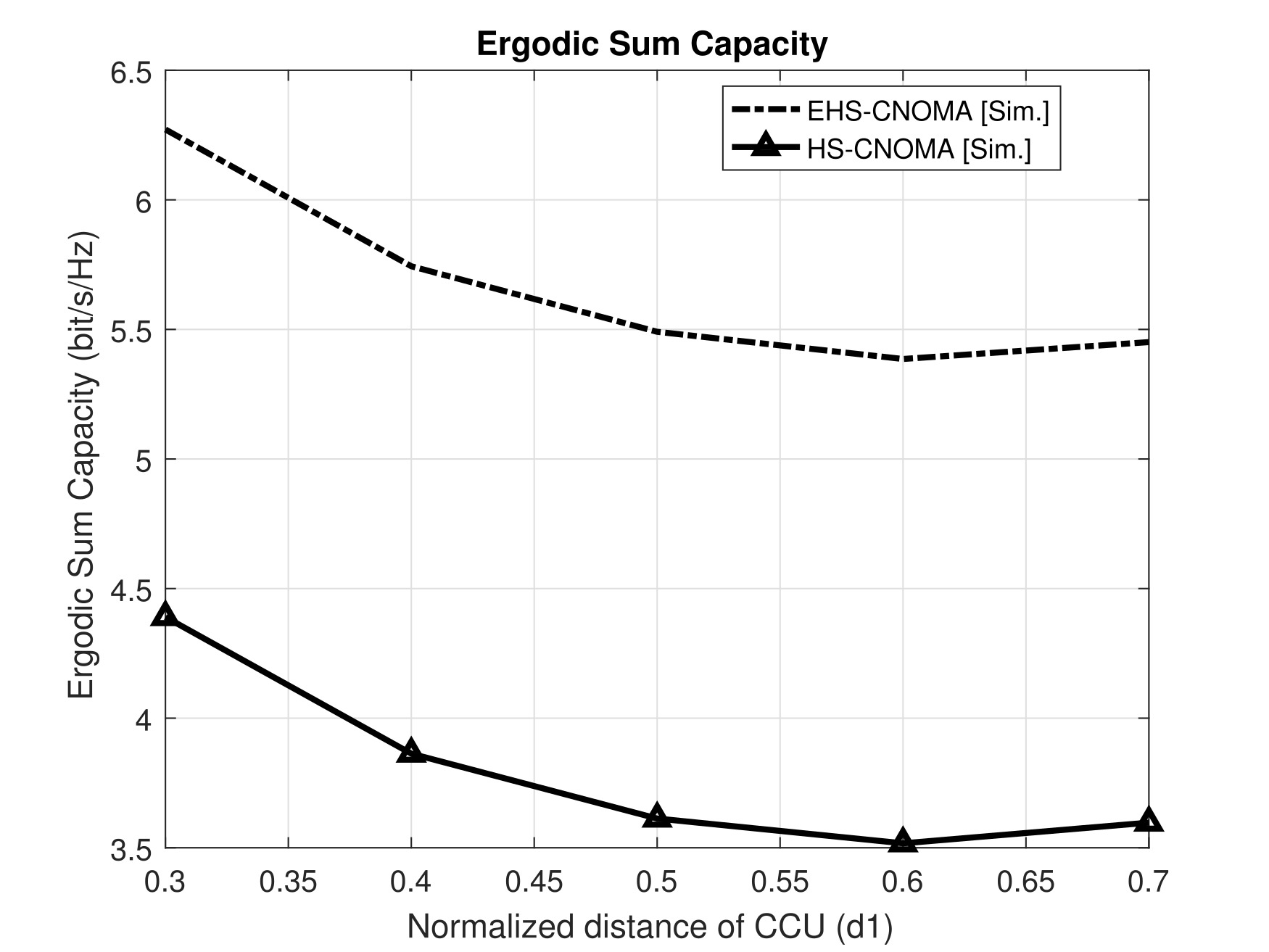}
\caption{ ESC versus $d_1$ of EHS-CNOMA and HS-CNOMA}
\label{image-myimage}
\end{figure}

\begin{figure}[h!]
\centering
\includegraphics[width=0.65\textwidth]{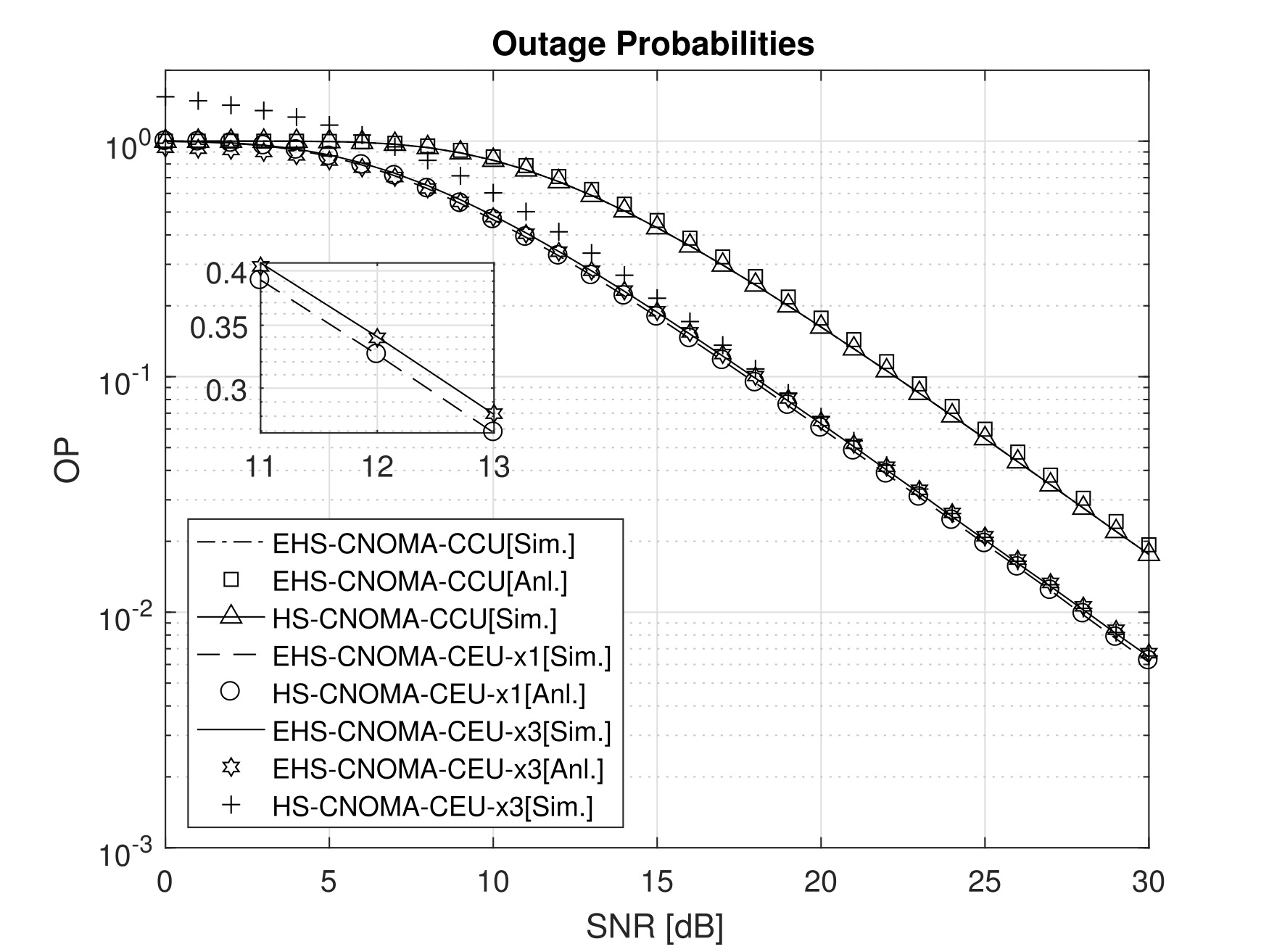}
\caption{ OP versus SNR performance of EHS-CNOMA and HS-CNOMA protocols}
\label{image-myimage}
\end{figure}

\begin{figure}[!t]
\centering
\includegraphics[width=0.65\textwidth]{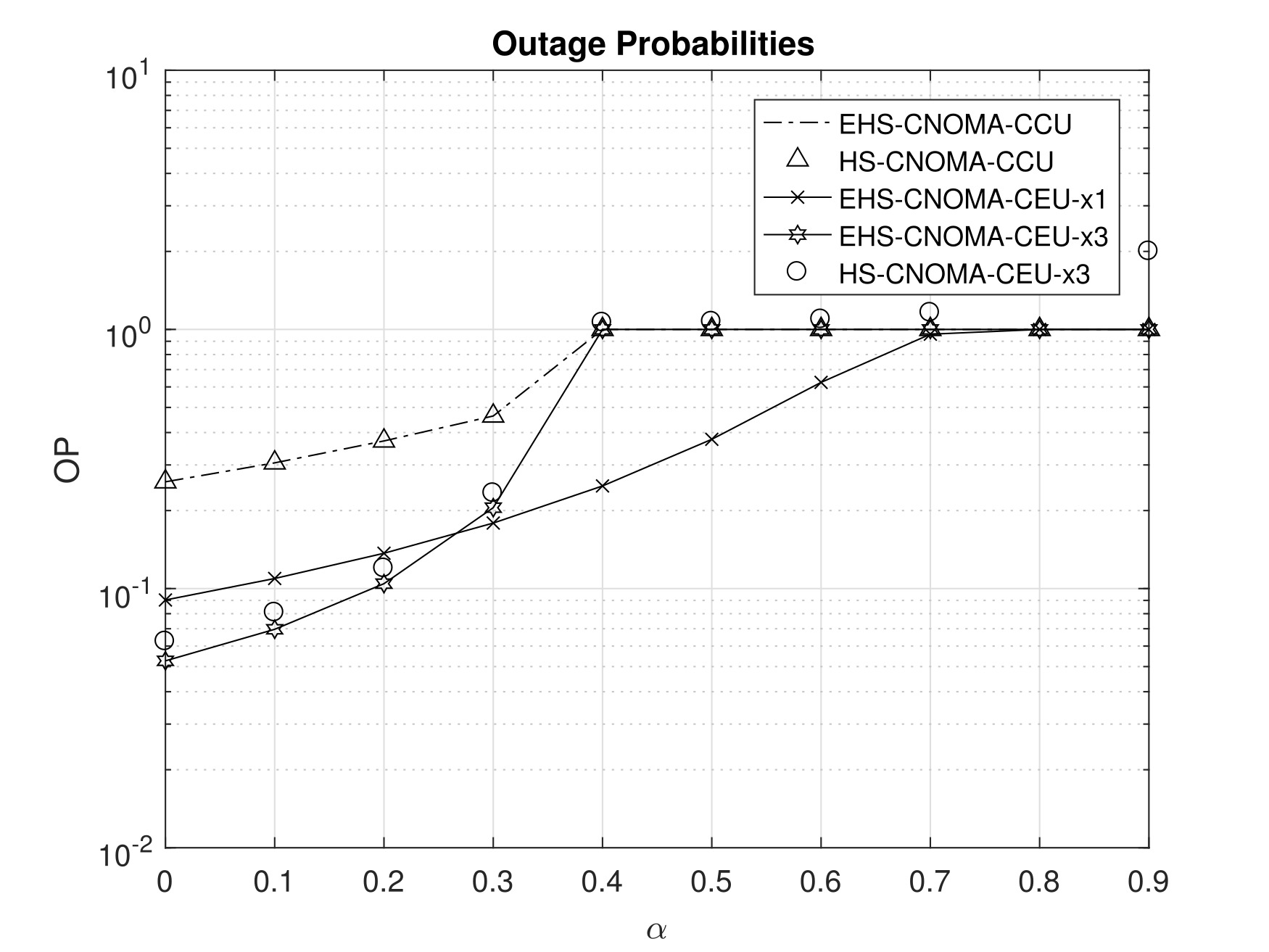}
\caption{ OP versus $\alpha$ of EHS-CNOMA and HS-CNOMA protocols}
\label{image-myimage}
\end{figure}

\begin{figure}[!t]
\centering
\includegraphics[width=0.65\textwidth]{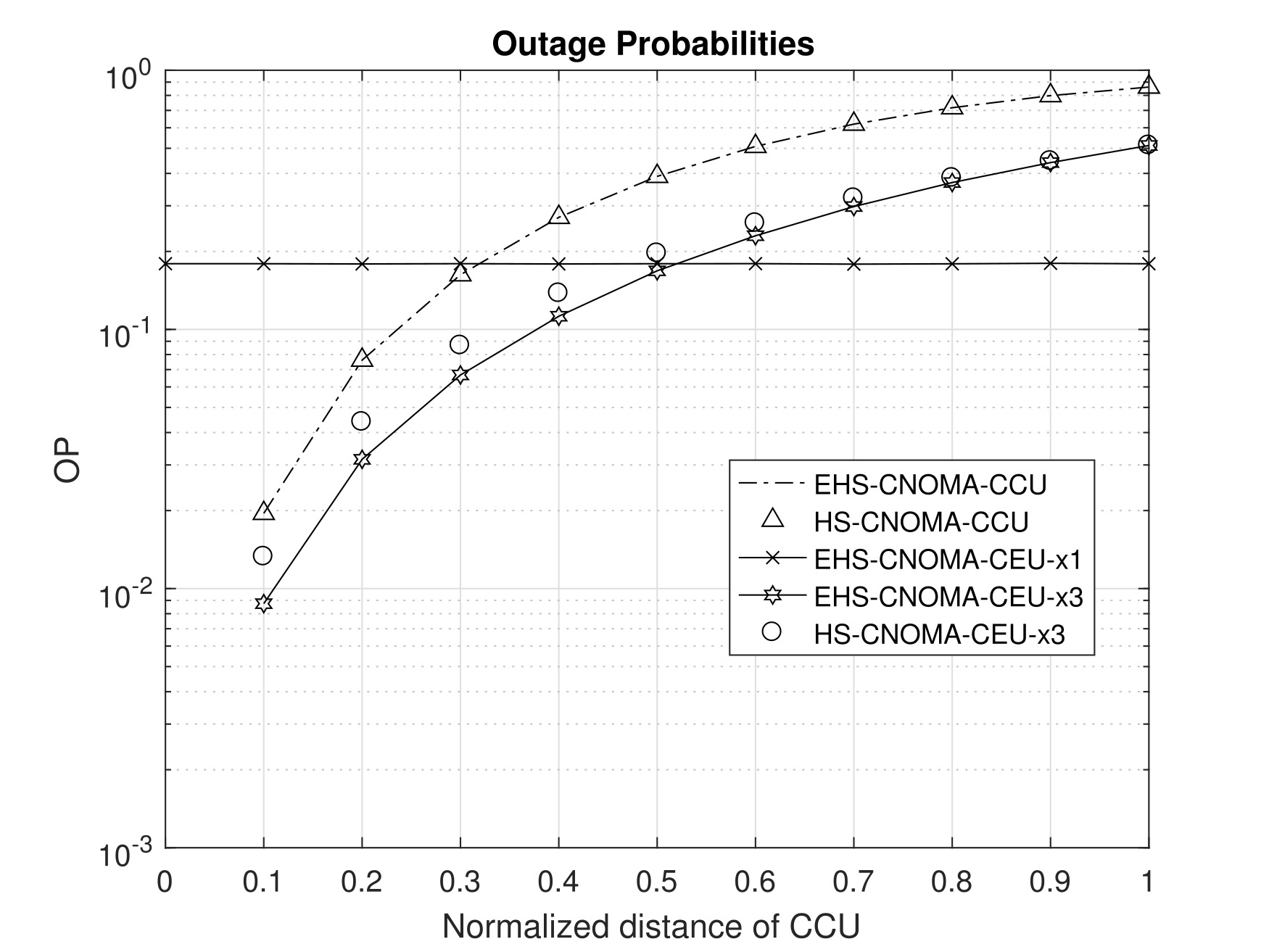}
\caption{ OP versus $d_1$ of EHS-CNOMA and HS-CNOMA protocols}
\label{image-myimage}
\end{figure}

\begin{figure}[!t]
\centering
\includegraphics[width=0.65\textwidth]{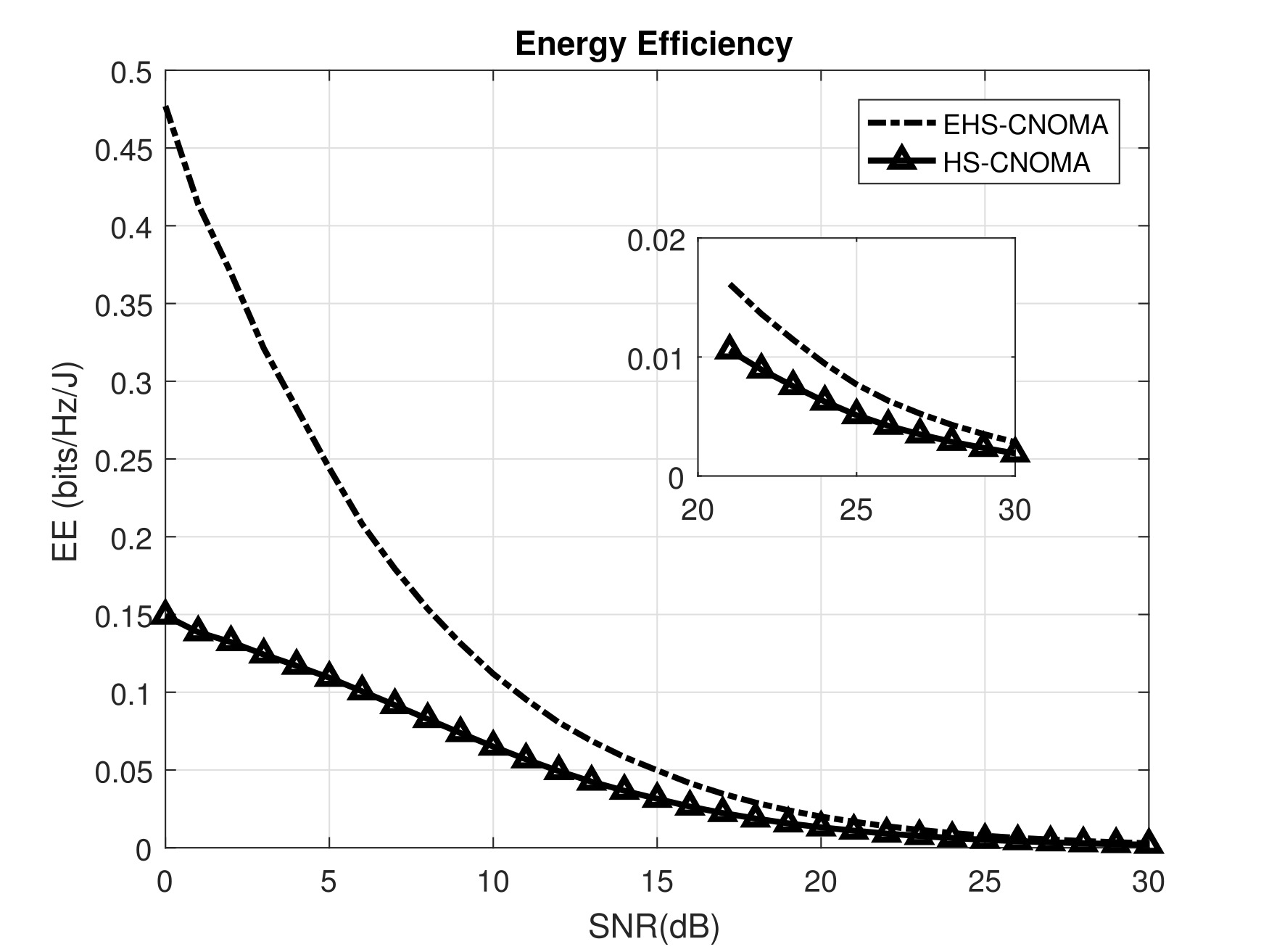}
\caption{ EE versus SNR performance of EHS-CNOMA and HS-CNOMA protocols}
\label{image-myimage}
\end{figure}

In the result analysis, the lines represent the simulation result and the markers represent analytical results respectively. For simulation purpose $\delta=0.3$ and $\alpha=0.3$ are considered [10-11,18]. Moreover, $\eta=0.7$, path loss exponent, $v=2$ and $R_{1}=R_2=R_3=1 bps/Hz$ are considered here for simulation purpose as well [2-4]. Additionally, $d_{1}=0.5$, $d_{2}=1$, $p_{N} = 0.1$ and  $p_{F} = 0.9$ are also considered for simulation purpose. The collinear outcomes of simulation and analytical results verify the appropriateness of the analysis for both ESC and OP. 

\subsection{Ergodic sum capacity}
 Fig.3 illustrates that EHS-CNOMA with MRC outplayed HS protocol in case of ESC by transmitting $x_1$ for the EH at CCU and information transfer for CEU. Higher SNR provides better ESC than lower SNR for all cases. Moreover, the analytical result also validates the simulation result as well for the EHS-CNOMA with MRC. 
\par
Fig.4 demonstrates the effect of the fraction of time block for $x_1$ ($\alpha$) on ESC. Moreover, $\rho=15 dB$ is considered in this case. So the EHS-CNOMA with MRC provides linearly increased ESC for $\alpha<0.7$. However, the provided ESC for EHS-CNOMA with MRC is saturated for $\alpha>=0.7$. Because for a longer duration of $\alpha T$, $(1-\alpha)T/2$ duration is shorter. So CCU and CEU cannot properly receive and decode the symbols properly for EHS-CNOMA with MRC. Though $x_1$ is transmitted successfully from BS to CEU for the proposed protocol hence it provides significantly higher ESC than others in case of considered CNOMA model. Moreover, the ESC of the HS-CNOMA with SC [10-11] is degraded due to the idle link is not utilized for $x_1$. In addition, CCU and CEU cannot receive and decode symbols ($x_2 \& x_3$) properly due to very short duration time for decoding. In this paper, ESC is compared with respect to $\alpha$ only because this is the dominating factor for HS protocol according to [18]. 
\par
Fig. 5 shows the comparisons between three cases for ESC with respect to $d_1$ in this case. Moreover, $\rho=15 dB$ is considered in this case as well. So this is clearly visible that for different values of $d_1$, EHS-CNOMA with MRC provides significantly higher ESC than other cases. The enhancement of ESC for the proposed EHS protocol is provided due to the utilization of unused link for $x_1$ and effective decoding by utilizing MRC at CEU for $x_3$. However, for the increasing distance between BS and CCU causes degradation of channel condition among them hence the ESC is decreased for all cases.   

\subsection{Outage Probability}

Fig. 6 illustrates that the EHS-CNOMA with MRC provides the same OP than HS-CNOMA with SC for $x_2$ at CCU. In case of the OP of $x_1$ for CEU, $x_1$ is only transferred with full power to CEU for the EHS protocol. So it is only compared with the analytical result of itself. Moreover, the link for $x_1$ is idle for the existing HS protocol, so the OP among two different cases cannot be compared. The OP of $x_3$ at CEU, EHS protocol with CNOMA provides lower OP than HS-CNOMA with SC for lower SNR due to MRC. The analytical results also validate the simulation results as well for the proposed EHS protocol with the considered model. All OP for CCU and CEU are decreased linearly along with the increasing SNR. 
\par
Fig.7 demonstrates the effect of the fraction of time block for $x_1$ ($\alpha$) on OP of CCU and CEU for different symbols. Moreover, $\rho=15 dB$ is considered in this case. Fig 8. depicted that for increasing values of $\alpha$ decreased the $(1-\alpha)T/2$ time duration. That is why CCU and CEU cannot receive and decode their corresponding symbols appropriately. As a result, OP is increased linearly for higher values of $\alpha$. Fig.8 also shows that the OP of $x_2$ for CCU increased linearly for an increasing number of $\alpha$ for all protocols. According to the EHS protocol, $x_1$ is transmitted directly to CEU from BS with full power. But $x_1$ cannot receive and decode properly at CEU due to the extremely higher time duration of $x_1$. Moreover, higher values of $\alpha$ also reduce the other time duration of EHS protocol significantly. So the OP of $x_1$ at CEU is increased linearly for higher values of $\alpha$. In case of the OP of $x_3$ at CEU, EHS-CNOMA with MRC provides less OP than HS-CNOMA with SC due to MRC. Though the OP for both cases increases to a great extent for $\alpha>=0.4$. Because lower time duration of decoding and relaying caused the huge outage event for $x_3$ at CEU.    

Fig. 8 shows the effect of $d_1$ on OP of CCU and CEU for different symbols. Moreover, $\rho=15 dB$ is considered in this case as well. So this is clearly visible that for different values of $d_1$, EHS-CNOMA with MRC provides the same OP as HS-CNOMA with SC for $x_2$ at CCU. Though the OP is increased for $x_2$ at CCU because the distance between BS and CCU is increased. Hence the channel condition among them has degraded accordingly. So it cannot decode $x_2$ successfully at CCU. For the OP of $x_3$ at CEU, the EHS-CNOMA with MRC provides lower OP than HS-CNOMA with SC for MRC at CEU. However, for higher $d_1$, the OP of $x_3$ at CEU increased linearly due to the channel condition between BS and CCU. The EHS-CNOMA with MRC provides the constant OP for $x_1$ at CEU. Because $x_1$ is transmitted directly from BS to CEU with full power. Only $x_3$ is relaying from CCU to CEU for all cases.

\subsection{Energy Efficiency}

Fig.9 illustrates the EE comparisons between the proposed EHS-CNOMA scheme and HS-CNOMA scheme as well. As EE is related to the ESC and $P_{CCU}$. The EHS protocol provides higher ESC than conventional HS protocol which is illustrated in Fig. 3. Hence, the EE is significantly higher for the proposed EHS-CNOMA compared to HS-CNOMA in Fig.9. However, the EE is comparatively much higher for the less SNR but EE is decreased for higher SNR in case of both cases. Because due to the increasing values of ESC with respect to SNR, the $P_{CCU}$ is also increased for both cases based on the harvested energy. Moreover, the difference between the achieved EE between EHS-CNOMA and HS-CNOMA is significantly higher for lower SNR compared to higher SNR.

\section{Conclusion}\label{sec10}

In this paper, the performance of EHS-CNOMA has been analyzed, where CCU is used as a relay for CEU. The ESC and OP of the proposed EHS-CNOMA with MRC technique are also investigated with their analytical derivations. The analytic results have been validated by the simulation results. Moreover, the advantages of the proposed EHS protocol have been demonstrated by numerical result analysis. To evaluate the system performance, the impact of different parameters on ESC and OP has been investigated and compared with HS-CNOMA. Moreover, the EE of the EHS-CNOMA is analyzed and compared with HS-CNOMA as well. The proposed protocol provides a significantly higher capacity than other protocols. Moreover, the considered CNOMA strategy also improved the OP of CCU and CEU for different symbols. Furthermore, EE for the proposed EHS-CNOMA is significantly higher than HS-CNOMA as well. However, the decoding complexity and SIC complexity in the considered model is same as existing CNOMA [2-6]. In the future, the work can be extended by integrating the EHS with amplify and forward relay assisted CNOMA [3].

\section{Acknowledgments}\label{sec11}
This work was supported by Priority Research Centers Program through the National Research Foundation of Korea (NRF) funded by the Ministry of Education, Science and Technology (2018R1A6A1A03024003).

\bibliographystyle{unsrt}  


\end{document}